\documentclass[prc,preprint,showpacs]{revtex4}

\begin{document}

\title{Comment on ``Interaction induced deformation of the momentum 
  distribution of spin polarized nuclear matter'' by Frick, M\"uther, and Sedrakian}

\author{S. Frauendorf}
  \affiliation{Department of Physics, University of Notre Dame,
Notre Dame, IN 46556, USA}
  \email{sfrauend@nd.edu}

\author{K. Neerg\aa rd}
  \affiliation{N\ae stved Gymnasium og HF, Nyg\aa rdsvej 43,
    DK-4700 N\ae stved, Denmark}
  \email{neergard@inet.uni2.dk}

\begin{abstract}

The paper 
does not take into account previous work where the effect they discuss is explored in
detail. It is known from this previous work that the approach is incomplete.

\end{abstract}


\maketitle

In the recent paper ``Interaction induced deformation of the momentum 
  distribution of spin polarized nuclear matter''~\cite{fms02}, 
Frick, M\"uther, and Sedrakian discuss the distribution of the
single-nucleon momentum in spin polarized nuclear matter. They find that
for a given direction of the nucleonic spin this distribution is
anisotropic, and estimate the gain in energy due to the deviation from
an isotropic distribution. They ascribe this effect to the exchange part
of a term in the tensor component of the one-pion exchange potential. As
mentioned in Ref.~\cite{fms02} the bare two-nucleon interaction has in
fact non-central terms, some of which may be expressed in the form of a
tensor force. A part of the tensor force originates in one-pion
exchange. The non-central terms in the bare interaction give rise to
similar terms in the in-medium effective interaction. In
Ref.~\cite{fms02} it is not mentioned that already in the literature of
about a quarter of a century ago it was pointed out that the exchange
part of the effective tensor force produces in spin polarized nuclear
matter an anisotropic distribution of the momenta of the nucleons with a
given direction of spin. Moreover, for almost that
long time the approach of   Ref.~\cite{fms02} has been known to
be incomplete.

To our knowledge the observation that in spin polarized nuclear matter
the exchange part of the tensor force produces a spin-dependent
single-nucleon potential which varies with the direction of the
nucleonic momentum was first made around 1975 in papers by Dabrowski and
Haensel~\cite{dh74} and Haensel~\cite{h75}. This observation leads
immediately to the conclusion that the Fermi surface must exhibit a
spin-dependent deviation from sphericity. The structure of the Fermi
surface of spin polarized nuclear matter was  investigated in
detail in subsequent papers by Haensel and Dabrowski~\cite{hd75} and
Dabrowski and Haensel~\cite{dh76}. The contributions of the distortions
of the Fermi surface to the spin and spin-isospin symmetry energy
coefficients and such quantities as the magnetic susceptibity of neutron
matter are also calculated in these papers. The framework of these
investigations is an extended Landau liquid model with a quasiparticle
interaction derived from Brueckner theory. The spin degree of freedom is
treated af if the nucleons with different spin along a fixed axis formed
two separate interacting fluids. This is also the way the nucleonic spin
is treated in Ref.~\cite{fms02}.

With the rise of the field of high spin nuclear physics around 1975,  
we began to study the system of rotating infinite nuclear matter. Since
angular momentum may be formed by alignment of the nucleonic spin, it
was neccessary to consider the influence of a spin
polarization. Due to unfortunate personal reasons, the results were finally published 
only in 1996 in Ref. ~\cite{fn96}.   A brief
account of our results~\cite{fn80} was prepared for a conference in
1980, and in 1981 a preliminary version of Ref.~\cite{fn96} was
distributed as a preprint~\cite{fn81}. We found that for some directions of the nucleonic
momentum the spin dependent field is quantized in a direction different
from that of the bulk polarization. Thus the approach taken in
Refs.~\cite{dh74,h75,hd75,dh76} is incomplete. We also found that to
first order in the polarization the distortion of the Fermi surface is
given entirely by the sum of a monopole and a quadrupole term. 
 A theory of non-rotating nuclear
matter in a spin-dependent external field such as neutron matter in a
 magnetic field is
obtained by setting in Eqs.~(55) of Ref.~\cite{fn96} $\omega=1$, and in
Eqs.~(56) $F(m_t)$ equal to minus the energy in the external field of a
nucleon with spin and isosin $(s_z,t_z)=(\frac12,m_t)$, and $G(m_t)=0$.
(Note that a factor $\hat k(m_t)$ is missing in the denominator in the
expressions for $A(m_t,\tilde m_t)$ and $B(m_t,\tilde m_t)$ in
Eqs.~(57).)

The incompleteness of the approach taken in
Refs.~\cite{dh74,h75,hd75,dh76} was pointed out in Ref.~\cite{fn81}. This
apparently inspired the authors of Refs.~\cite{dh74,h75,hd75,dh76} to
revisit the problem. In fact papers published in 1982 by Haensel and 
Jerzak~\cite{hj82} and in 1984 by Dabrowski~\cite{d84} cite 
Ref.~\cite{fn81}, and in Ref.~\cite{d84} the polarization induced 
distortion of the Fermi surface is illustrated graphically in a form 
also employed in Refs.~\cite{fn80,fn81}. (See Fig. 1 of Ref.~\cite{fn96} and 
Figs. 3 and 5 of Ref.~\cite{d84}.) The analysis in Refs.~\cite{hj82,d84} 
is complete.

\bibliography{frick}

\end{document}